\documentclass[11pt,twoside]{article}

\usepackage{asp2006}
\usepackage{fancyhdr,graphicx,caption,rotating,multirow,lscape}
\DeclareGraphicsExtensions{.eps,.eps.gz,.ps,.jpg,.tiff}

\begin{document}

\title{Transdet: a matched-filter based algorithm for transit detection --
application to simulated COROT light curves}  



\author{P. Bord\'e\altaffilmark{\dagger}}
\affil{Harvard-Smithsonian Center for Astrophysics, Cambridge, MA 02138, USA}
\altaffiltext{1}{Michelson Postdoctoral Fellow. Current address: Michelson Science Center, Caltech, Pasadena, CA 91125, USA.}
\author{F. Fressin}  
\affil{Gemini, Observatoire de la C\^ote d'Azur, 06304 Nice, France}    
\author{M. Ollivier and A. L\'eger}
\affil{IAS, Universit\'e Paris-Sud, 91405 Orsay, France}
\author{D. Rouan}
\affil{LESIA, Observatoire de Paris, 92195 Meudon, France} 

\begin{abstract} 
We present a matched-filter based algorithm for transit detection and its application to simulated COROT light curves. This algorithm stems from the work by Bord\'e, Rouan \& L\'eger (2003, A\&A 405, 1137). We describe the different steps we intend to take to discriminate between planets and stellar companions using the three photometric bands provided by COROT. These steps include the search for secondary transits, the search for ellipsoidal variability, and the study of transit chromaticity. We also discuss the performance of this approach in the context of blind tests organized inside the COROT exoplanet consortium.
\end{abstract}


%
%
\section{Introduction}
COROT \citep{Baglin2003}, scheduled for launch by the end of 2006, will be the first space mission with the capability to detect extrasolar planets with sizes down to a couple of Earth radii \citep{Rouan2000}. It will observe a total of 60,000 stars in 5 runs of 150 days each. In a previous paper, we worked out a theoretical estimate of the planet detection capability of COROT \citep{Borde2003}. In this paper, we first describe our transit detection algorithm, dubbed \emph{Transdet}, then we report its performance on simulated light curves (LCs), and finally, we discuss the tools we are developing to discriminate between genuine exoplanets and astrophysical false positives (binary stars).

%
%
\section{Key ideas about Transdet} \label{sec:algo}
At low signal-to-noise (SNR), a transit signal can be approximated by a rectangular signal oscillating between a high level $H$ (outside the transit) and a low level $L$ (during the transit). The amplitude of this signal depends on the ratio of the planet surface to that of the star: $H-L = \varepsilon H$, where $\varepsilon \equiv (R_p/R_\star)^2$. The time spent at level $L$, or transit duration $\tau$, is much shorter than the transit period $T$. We note $t_0$ the date of the beginning of the first transit falling inside the observation window (150 days for COROT).

In order to detect the transit signal in a noisy LC, we adopt the matched-filter approach that consists in correlating the LC with rectangular signals of the type described above. We call \emph{test-signals} these rectangular signals. Each of them is characterized by a parameter triplet $(\hat{T},\hat{\tau},\hat{t_0})$. The correlation reaches a peak when $(\hat{T},\hat{\tau},\hat{t_0}) = (T,\tau,t_0)$. Practically, LCs must be first high-pass filtered to remove irrelevant low-frequency variations (stellar fluctuations) that would bias the correlation, and we take the negative of the LCs so that transits would produce positive correlation peaks.

For, this method to work, the three-dimensional space $(\hat{T},\hat{\tau},\hat{t_0})$ must be explored with a sufficient resolution in the given range accessible to COROT ($1 \le \hat{T} \le 75\:\mathrm{days}$, $1 \le \hat{\tau} \le 10\:\mathrm{hrs}$, $0 \le \hat{t_0} \le T$). The choice of this resolution results from a trade-off between computing time and detection efficiency. Typically, we use steps $(\delta\hat{T},\delta\hat{\tau},\delta\hat{t_0})$ small enough so that the maximum correlation signal would be no less than 75\,\% of its theoretical maximum value. This requirement translates into the computation of $14\times10^6$ correlations per LC, a task completed in 90~s with a fully IDL-coded algorithm on a Pentium~M at 1.6~Ghz. However, preliminary tests indicate that an optimized IDL--C hybrid code could cut down the computing time by a factor 7--8, which would make possible a higher detection efficiency.

%
%
\section{Detection threshold} \label{sec:threshold}
%
%
\subsection{Theoretical detection threshold} \label{sub:theoretical}
Practically, we use as our primary detection metric not the correlation itself, but the SNR on the correlation defined as
\begin{equation} \label{eq:snr1}
S\!N\!R_1 \equiv (k\tau)^{1/2}\,\varepsilon H/\sigma,
\end{equation}
where $\sigma$ is the standard deviation of the noise affecting the high-pass filtered LC, and $k$ is the number of transits in the observation window.

A theoretical detection threshold can be simply set assuming: (1) a total number of LCs, (2) a false detection rate, and (3) Gaussian white noise \citep[see][p.\ 45]{BordePhD}. Requiring less than 1 false detection, we obtain $S\!N\!R_1 > 6.5$ for 1,000 LCs, and $S\!N\!R_1 > 7.0$ for 60,000 LCs.

%
%
\subsection{Empirical detection threshold} \label{sub:empirical}
In order to train the co-investigators of the COROT exoplanet consortium and compare the performances of existing transit detection algorithms, a first transit detection blind test (BT1) was organized inside the consortium in 2004 \citep{Moutou2005}. BT1 consisted in the analysis of 999 LCs with no information on the target stars.

\begin{figure}[h]
\centering
\begin{minipage}[c]{.49\textwidth}
\includegraphics[width=6.5cm]{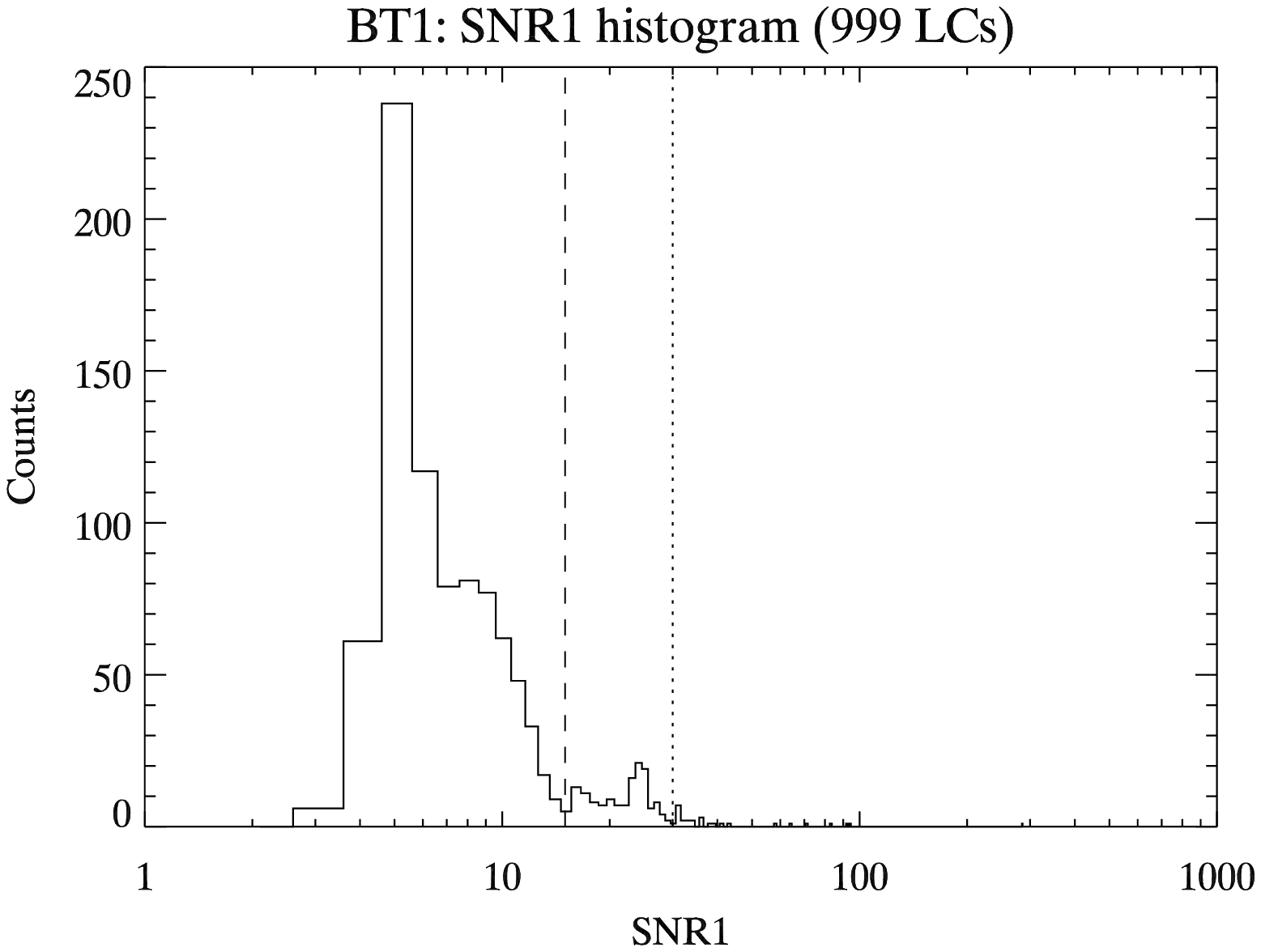}
\end{minipage}
\begin{minipage}[c]{.49\textwidth}
\includegraphics[width=6.5cm]{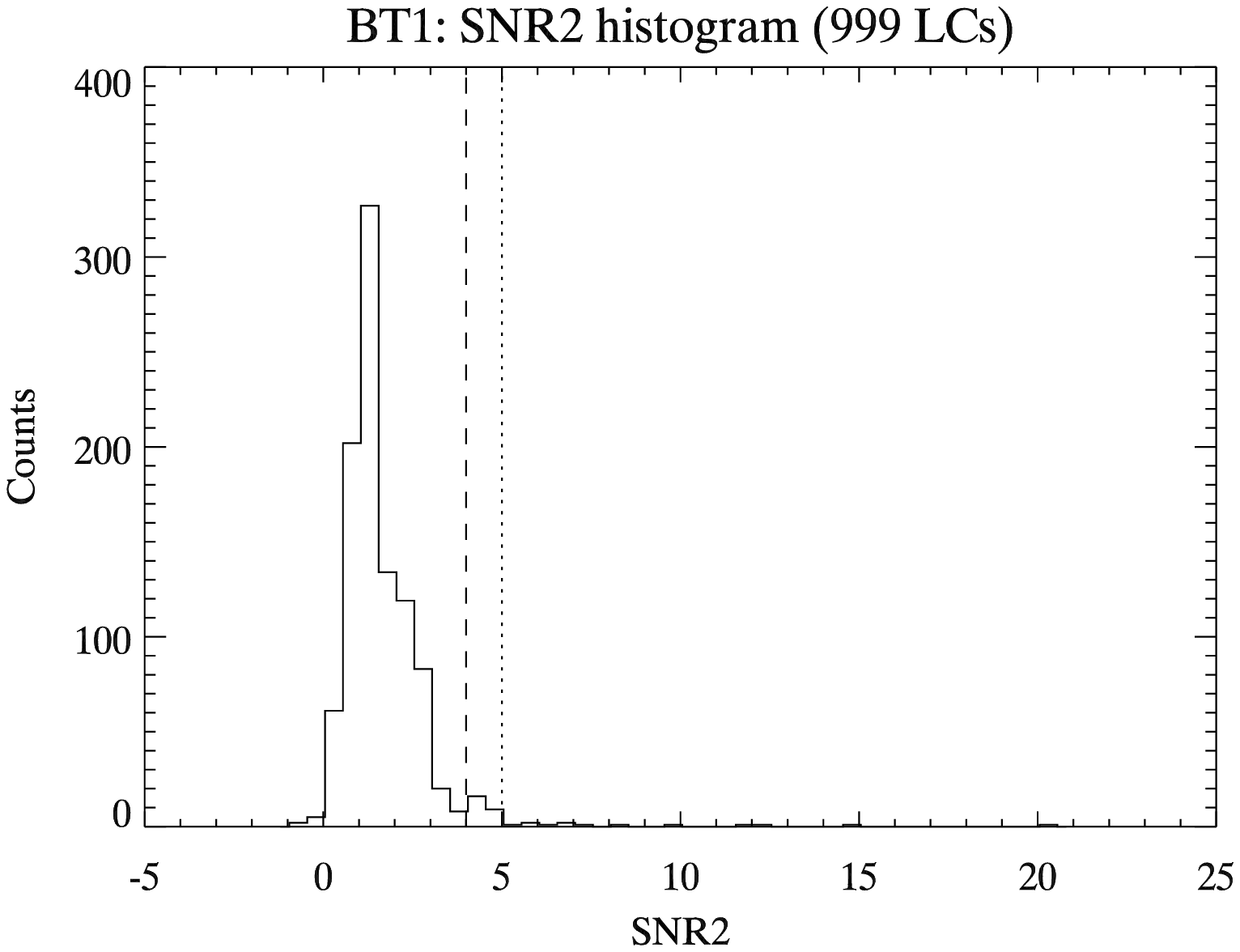}
\end{minipage}
\caption{Left: $S\!N\!R_1$ histogram. The dashed and dotted lines mark thresholds at 15 and 30, respectively. Right: $S\!N\!R_2$ histogram. The dashed and dotted lines mark thresholds at 4 and 5, respectively. \label{fig:histograms}}
\end{figure}

Figure~\ref{fig:histograms} (left) shows the distribution of $S\!N\!R_1$ for BT1. A priori, one would expect a bimodal distribution, where LCs without transits would appear at low $S\!N\!R_1$, whereas LCs with transits would appear at high $S\!N\!R_1$, and a dividing line around 6.5. If the obtained distribution is indeed bimodal, the plot suggests a dividing line, or detection threshold, of either $\approx 15$ (175 selected LCs) or $\approx 30$ (28 selected LCs), largely above the theoretical estimate. This discrepancy stems from the inaccuracy of the white noise hypothesis as pointed out by \citet{Pont2006}.

\begin{figure}[!ht]
\centering
\includegraphics[width=9cm]{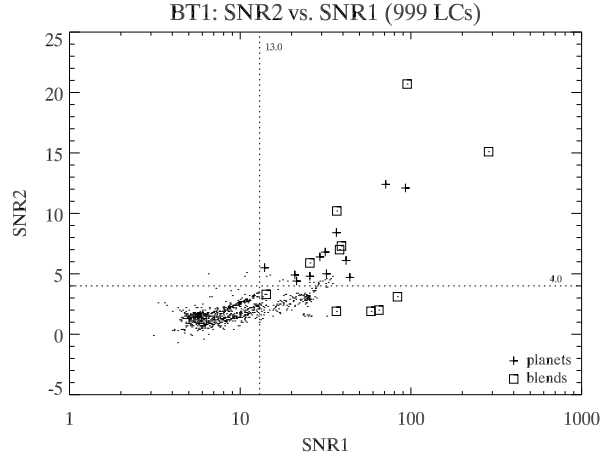}
\caption{$S\!N\!R_2$ vs. $S\!N\!R_1$ diagram. Crosses and squares are overplotted on genuine planets and binary stars, respectively. A thresholding in $S\!N\!R_2$ selects all planets much more efficiently than a thresholding in $S\!N\!R_1$. \label{fig:snr1snr2}}
\end{figure}

We find that a secondary detection metric computed on the folded LC (FLC) helps the selection process. We define this new metric as
\begin{equation} \label{eq:snr2}
S\!N\!R_2 \equiv (H-L)/\sigma_f,
\end{equation}
where $H$ and $L$ are measured on the FLC, and $\sigma_f$ is the standard deviation of the noise affecting the high-pass filtered FLC. Figure~\ref{fig:histograms} (right) shows the distribution of $S\!N\!R_2$ for BT1. This new plot suggests a threshold of either $\approx 4$ (42 selected LCs) or $\approx 5$ (16 selected LCs). It turns out that $S\!N\!R_2 > 4.0$ is the right criterion as the 12 genuine planets correctly detected by \emph{Transdet} in this exercise are among the 42 selected LCs (see also Fig.~\ref{fig:snr1snr2}).

%
%
\subsection{Detection capability} \label{sub:capability}
\begin{figure}[!ht]
\centering
\includegraphics[width=9cm]{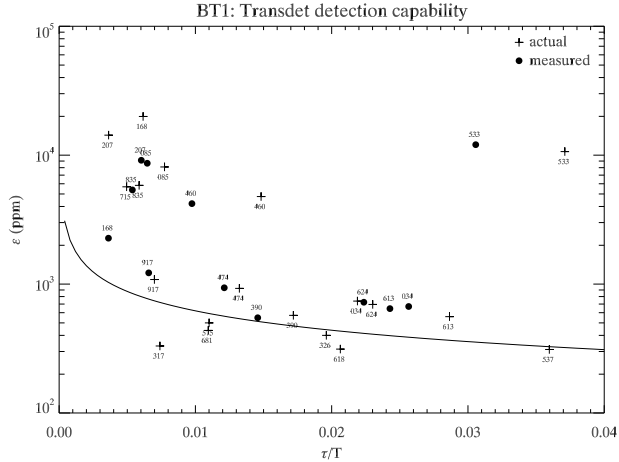}
\caption{Comparison between actual and measured values for $\varepsilon$ and $\tau/T$ for BT1. A line at constant $S\!N\!R_1$ marks the bottom of the region where transits were detected.\label{fig:capability}}
\end{figure}

Figure~\ref{fig:capability} is an attempt to estimate the detection capability of \emph{Transdet} from the results of BT1 in much the same way as in \citet[Fig.\ 7]{Moutou2005}. The solid line is a line at constant $S\!N\!R_1$ adjusted so that all detected planets lie above it, and all undetected planets lie under it. From this, we conclude that the detection limit would be about 300~ppm at $\tau/T = 0.04$, which could be a 2 Earth-radii planet around a G2V star with a 3-day period.

%
%
\section{False positive discrimination} \label{sec:false}
The vast majority of transit signals are not caused by planets, but rather by stellar companions in binary stars \citep[e.g.,][]{Brown2003}: grazing eclipsing binary (GEB) or background eclipsing binary (BEB) blended with foreground stars may produce transits with amplitudes consistent with planetary companions. Radial velocimetry provides the definitive way to discriminate planets from binaries. However, it would be much too time-consuming to observe every single candidate, thus it is mandatory for the success of transits searches to sort out as many candidates as possible using the photometry alone. A second transit detection blind test (BT2) was organized inside the COROT consortium in 2006 to work toward this goal. BT2 participants were given 237 LCs measured in the three COROT photometric channels, as well as the spectral types of the target stars.
%
%
\subsection{Anomalies on folded light curve}
A first clear indicator of the stellar nature of the companion is the presence of a secondary eclipse in the LC. Therefore, once the primary eclipse has been found, we scan the FLC for shallower transit-like events.

A second indicator is the presence of ellipsoidal variability: tidal forces exerted by a massive companion at short orbital distance distort the primary's shape into an ellipsoid. Because the ellipsoid rotates with the orbital motion, the apparent stellar disk varies and the LC's baseline is modulated at twice the orbital frequency \citep[see][]{Sirko2003}. We also scan the FLC for this modulation.

%
%
\subsection{Transit parameter check}
COROT's entry catalog contains photometrically-determined spectral types for target stars, as well as estimates for stellar masses and radii. With the target radius and the measured transit depth, it is straightforward to check that the companion's radius lies in the planet range, say $R_p \le 1.5\:\mathrm{R_J}$.

For short orbital periods where the assumption of a circular orbit is reasonable, the transit duration can be checked against the maximum transit duration computed for an equatorial transit, i.e., $\tau_\mathrm{max} = 1.8\,R_\star (T/M_\star)^{1/3}$, where $\tau_\mathrm{max}$ is in hours, $T$ in days, and $R_\star$ and $M_\star$ in solar units. A measured duration greater than $\tau_\mathrm{max}$ would either point to a problem with the transit parameters (e.g., an incorrect period) or with the parameters of the star in front of which the transit is occurring, thus betraying a BEB with a spectral type different than that of the target.

Finally, the transit shape can be an indicator of GBs for partial occultations produce V-shape transits. To quantify this, we fit a trapeze to the transit in order to measure the duration $\tau_F$ of the flat part of the transit (between ingress and egress) as well as the total transit duration $\tau$, and we compute $V\!S \equiv 1-\tau_F/\tau$. Currently, our V-shape criterion is arbitrarily defined as $V\!S > 0.8$.

%
%
\subsection{Transit chromaticity}
\begin{figure}[!ht]
\centering
\includegraphics[width=9cm]{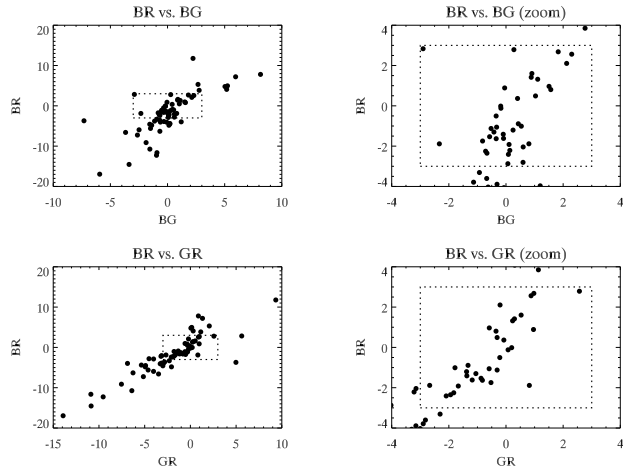}
\caption{Chromaticity diagrams for a selection of BT2 LCs. Dotted boxes mark the $3\,\sigma$ limit.\label{fig:color}}
\end{figure}

COROT will have the unique capability to measure transits in three spatially-separated color channels thanks to a prism objective. Only BEBs perfectly aligned with target stars will produce transits in the three channels. Therefore, transit chromaticity will be a powerful indicator of a false positive. For the sake of simplicity, COROT's color channels will be referred to as $B$ (blue), $G$ (green), and $R$ (red), but note that they do not correspond to any existing photometric systems. We define now three chromaticity indicators,
\begin{equation}
BG \equiv \frac{\varepsilon_B-\varepsilon_G}{(\sigma_B^2+\sigma_G^2)^{1/2}} \quad
BR \equiv \frac{\varepsilon_B-\varepsilon_R}{(\sigma_B^2+\sigma_R^2)^{1/2}} \quad
GR \equiv \frac{\varepsilon_G-\varepsilon_R}{(\sigma_G^2+\sigma_R^2)^{1/2}},
\end{equation}
where $\varepsilon_B$, $\varepsilon_G$, and $\varepsilon_R$ are the relative transits depths in the color channels, and $\sigma_B$, $\sigma_G$, and $\sigma_R$, the corresponding standard deviations of the noise. Figure~\ref{fig:color} shows chromaticity diagrams for a selection of BT2 LCs. In this case, we applied a $3\,\sigma$ threshold to classify LCs as chromatic.

%
%
\section{Conclusion}
We have developed \emph{Transdet}, a matched-filter based algorithm for transit detection, as well as tools for false positive discrimination using the LCs alone. All of these were developed and tested using blind tests conducted within the COROT consortium. However, fine tuning of the thresholds for the various tests will be done by studying the statistics of real COROT data.

\acknowledgements 
This work was performed in part under contract 1256791 with the Jet Propulsion Laboratory (JPL), funded by NASA through the Michelson Fellowship Program. JPL is managed for NASA by the California Institute of Technology. This research has made use of NASA's Astrophysics Data System.


\end{document}